\documentclass[conference]{IEEEtran}
\IEEEoverridecommandlockouts
\usepackage{cite}
\usepackage{amsmath,amssymb,amsfonts}
\usepackage{graphicx}
\usepackage{textcomp}
\usepackage{xcolor}
\usepackage{multirow}
\usepackage{float}
\usepackage{stfloats}
\usepackage{enumitem}
\usepackage{hyperref}

\begin{document}

\title{Physically-Constrained Harmonic Separation for Robust Heart and Respiratory Rate Estimation from Wrist Photoplethysmography
\vspace{-0.2em}
\thanks{This research was partially funded by Mohammed VI Polytechnic University through the i-Respire research project.}
\thanks{\copyright~2026 IEEE. Personal use of this material is permitted. Permission from IEEE must be obtained for all other uses, in any current or future media, including reprinting/republishing this material for advertising or promotional purposes, creating new collective works, for resale or redistribution to servers or lists, or reuse of any copyrighted component of this work in other works.}
}

\author{

\IEEEauthorblockN{Nouhaila Fraihi\textsuperscript{1}, Ouassim Karrakchou\textsuperscript{1}, and Mounir Ghogho\textsuperscript{2,3}}
\IEEEauthorblockA{
\textsuperscript{1}TICLab, International University of Rabat, Morocco\\
\textsuperscript{2}College of Computing, Mohammed VI Polytechnic University, Morocco\\
\textsuperscript{3}Faculty of Engineering, University of Leeds, UK\\
Emails: \{nouhaila.fraihi, ouassim.karrakchou\}@uir.ac.ma, mounir.ghogho@um6p.ma}
\vspace{-2em}
}

\maketitle


\begin{abstract}
Wrist-worn photoplethysmography (PPG) enables continuous monitoring of cardiopulmonary physiology, but reliable heart rate (HR) and respiratory rate (RR) estimation in free-living conditions remains challenging due to non-stationary motion artifacts that spectrally overlap with physiological dynamics. Existing signal-processing methods degrade under strong motion, while unconstrained deep learning approaches often lack physiological interpretability and identifiable structure. We propose a Physically-Constrained Harmonic Separation (PCHS) framework that formulates HR and RR estimation from wrist PPG as an analysis-by-synthesis problem, where accelerometer measurements condition artifact separation rather than directly regressing vital signs. A physics-guided harmonic generator decomposes the observed signal into quasi-periodic physiological components and a motion-related residual, enabling HR recovery from the fundamental frequency and RR prediction from respiratory-driven modulations of the harmonic parameters. Robust reconstruction objectives, separation constraints, and uncertainty-aware weighting stabilize the decomposition under motion. Experiments on the motion-intensive PPG-DaLiA dataset demonstrate that PCHS outperforms state-of-the-art methods while yielding interpretable signal decompositions that effectively disentangle physiological activity from motion artifacts.
\end{abstract}


\section{INTRODUCTION}

Photoplethysmography (PPG), widely used in wrist-worn wearables, enables continuous monitoring of cardiovascular and respiratory activity; however, reliable heart-rate (HR) and respiratory-rate (RR) estimation in free-living conditions remains challenging due to motion artifacts that obscure physiological information~\cite{charlton2022wearable,kim2023photoplethysmography}.

The wrist PPG signal consists of a pulsatile cardiac component, slow respiratory-driven baseline variations, and coupled amplitude and frequency modulations~\cite{allen2007photoplethysmography}. In free-living conditions, motion artifacts are typically additive, highly non-stationary, and spectrally overlapping with both cardiac ($\approx0.5$--$4$\ Hz) and respiratory ($<0.5$\ Hz) dynamics~\cite{Charlton2023}, rendering fixed filtering and simple noise suppression ineffective.

To mitigate motion corruption, prior work has relied on signal-processing pipelines that use inertial measurements as motion references for artifact suppression, followed by spectral rate estimation. Representative methods employ accelerometer-guided sparse spectral reconstruction or adaptive time–frequency filtering to attenuate motion-related components before extracting HR from dominant spectral peaks~\cite{zhang2014troika,wijshoff2016reduction,zhou2020heart}. Respiratory rate (RR) estimation typically exploits respiration-induced amplitude, baseline, or frequency modulations of the PPG waveform, sometimes combined with inertial sensing to suppress motion-correlated energy in the respiratory band~\cite{charlton2016assessment,jarchi2019estimation}. While effective under controlled conditions, these approaches assume separability between motion and physiological dynamics, an assumption that breaks down in free-living settings due to non-stationary and spectrally overlapping motion artifacts.

More recently, deep learning (DL) approaches have been proposed to jointly perform artifact suppression and vital-sign estimation from PPG within unified frameworks. Some DL methods primarily operate on raw PPG signals, formulating HR and RR estimation as end-to-end regression tasks~\cite{demireltemporal,bian2020respiratory,osathitporn2023rrwavenet}. Building on these PPG-only formulations, subsequent DL frameworks incorporate inertial measurements to explicitly account for motion effects. These approaches fuse PPG and acceleration signals in a supervised manner to improve robustness under movement~\cite{reiss2019deepppg,benfenati2025enhanceppg,kazemi2025respiration}.

Despite recent accuracy gains, most deep learning--based approaches still formulate vital-sign estimation from PPG as a black-box regression problem. Even when incorporating accelerometer signals through supervised fusion, such models offer limited physiological interpretability and may rely on spurious correlations between motion patterns (e.g., activity type, movement intensity, or periodic arm swing) and HR or RR. This does not guarantee recovery of underlying physiological dynamics and can promote shortcut learning from motion cues~\cite{naeini2023deep}. These limitations have motivated knowledge-informed learning approaches that integrate physiological constraints to improve robustness and interpretability. Recent methods, such as KID-PPG, incorporate explicit motion-aware processing and probabilistic inference to enhance heart-rate estimation under motion~\cite{kechris2024kid}. Such formulations typically adopt simplified coupling assumptions between inertial signals and photoplethysmography. While effective in many scenarios, ensuring fully identifiable and physiologically disentangled representations remains challenging in the presence of non-stationary and spectrally overlapping motion artifacts.

 These limitations motivate approaches that treat motion as an explanatory source of corruption rather than a predictive shortcut, and that enforce identifiable, physiology-driven modeling. Motivated by this perspective, we propose a \emph{Physically-Constrained Harmonic Separation} (PCHS) framework for robust HR and RR estimation from wrist PPG under motion. PCHS formulates denoising as an analysis-by-synthesis problem in which a deep learning model, conditioned on accelerometer data, predicts time-varying physiological parameters, such as the cardiac fundamental frequency and harmonic amplitudes, that synthesize a pulse-synchronous PPG component, while remaining content, including motion artifacts and unmodeled distortions, is captured by an additive residual and a time-varying reliability estimate. This formulation yields interpretable cardiac and respiratory representations.

The main contributions of this work are:
\begin{itemize}[leftmargin=1.em]
\item \textbf{Cross-modal motion-conditioning encoder:} We introduce a motion-aware dual-stream encoder in which accelerometer measurements condition PPG representations via feature-wise linear modulation (FiLM), enabling motion-informed artifact attribution without directly regressing vital signs.

\item \textbf{Physically constrained analysis-by-synthesis with robust objectives:} We formulate PPG decomposition as an analysis-by-synthesis problem that predicts time-varying physiological parameters and synthesizes a pulse-synchronous PPG component using a harmonic model. Robust reconstruction and separation objectives stabilize the decomposition under motion while isolating non-physiological content in an additive residual.

\item \textbf{Uncertainty-aware vital-sign inference:} We estimate HR from the recovered cardiac fundamental frequency and RR from physiologically meaningful low-frequency modulations, using a learned time-varying reliability signal to downweight motion-corrupted intervals during inference.
\end{itemize}

The remainder of this paper is organized as follows. Section~\ref{sec:Methods} presents the proposed framework, detailing the model architecture, physiological decomposition, and training objective. Section~\ref{sec:Experiments} describes the experimental protocol, including the dataset, ablation studies, and comparative results. Finally, Section~\ref{sec:Conclusion} concludes the paper and outlines directions for future work.

\section{Methods}
\label{sec:Methods}

\subsection{Overview and Design Principles}
\begin{figure*}
    \centering
    \includegraphics[width=1\linewidth]{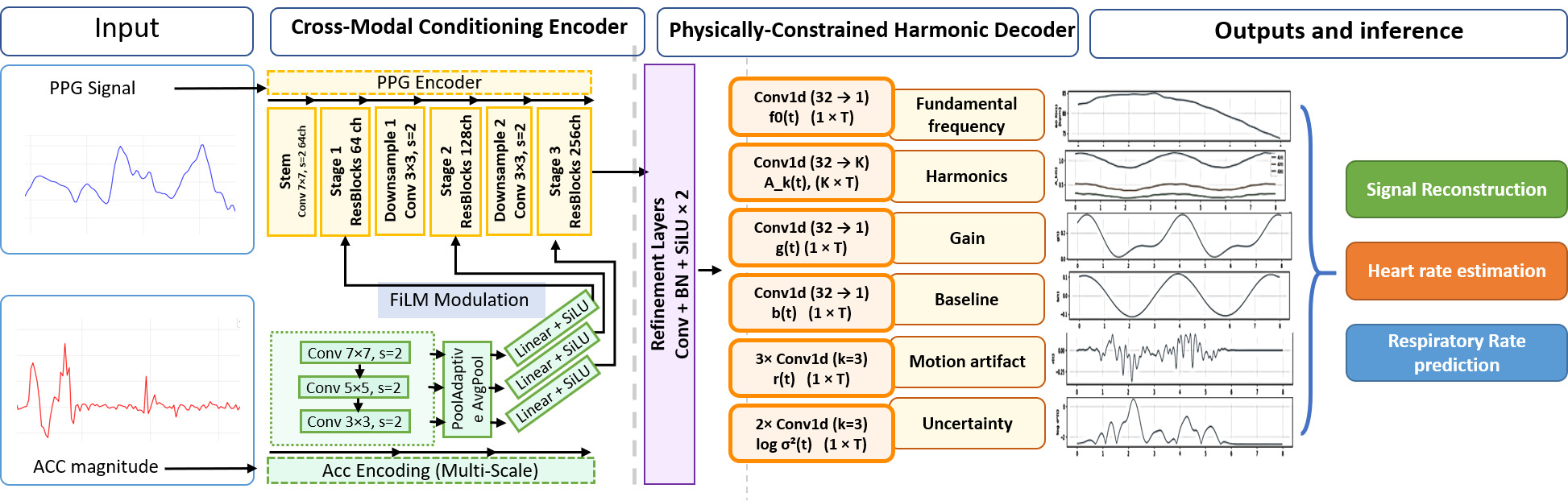}
    \caption{Overview of PCHS. A cross-modal encoder uses FiLM for motion adaptation, while a physics-constrained decoder synthesizes physiological parameters and residuals for reliability-aware inference.}
    \label{fig:placeholder}
\end{figure*}

We propose a \emph{Physically-Constrained Harmonic Separation} (PCHS) framework for robust HR and RR estimation from wrist-worn photoplethysmography (PPG) under motion. As illustrated in Fig.~\ref{fig:placeholder}, PCHS comprises three components:  
(i) a \emph{cross-modal motion-conditioning encoder}, in which a compact PPG encoder extracts multiscale, morphology-aware temporal features while a lightweight accelerometer (ACC) encoder captures motion; ACC features modulate PPG representations via feature-wise linear modulation (FiLM)~\cite{perez2018film} at multiple stages to enable motion-adaptive feature extraction;  
(ii) a \emph{physically constrained harmonic decoder}, which maps motion-conditioned features to time-varying physiological parameters, including cardiac fundamental frequency, harmonic amplitudes, gain, and baseline, synthesizing a pulse-synchronous physiological waveform via a harmonic model, while a residual branch captures non-harmonic, motion-consistent artifacts and an uncertainty head predicts local signal reliability; and  
(iii) an \emph{outputs and inference module}, which performs reliability-aware vital-sign estimation, inferring HR from the recovered fundamental frequency and predicting RR from physiologically meaningful parameter modulations, with uncertainty used to downweight motion-corrupted intervals. Overall, PCHS formulates signal decomposition and vital-sign estimation as an \emph{analysis-by-synthesis} problem, in which physiological structure is enforced by construction through a harmonic generative model, improving identifiability, interpretability, and robustness in free-living conditions.

\subsection{Problem Formulation}
Let $x(t)\in\mathbb{R}$ denote a wrist-worn PPG measurement at time $t/f_s$, with $f_s$ being the sampling frequency, and let $\mathbf{u}(t)\in\mathbb{R}^{3}$ denote the synchronized tri-axial ACC signal. To obtain a rotation-invariant motion representation, we compute the accelerometer magnitude
\begin{equation}
u(t)=\|\mathbf{u}(t)\|_2.
\end{equation}
Following prior wrist-PPG formulations that model motion artifacts as non-stationary additive components~\cite{jarchi2019estimation,kechris2024kid}, we express the observed signal as
\begin{equation}
x(t)=\mu_{\mathrm{phys}}(t)+r(t)+\epsilon(t), \qquad t=1,\cdots,T
\label{eq:obs}
\end{equation}
where $\mu_{\mathrm{phys}}(t)$ denotes the clean, quasi-periodic physiological component dominated by cardiac and respiratory activity, $r(t)$ represents motion-induced artifacts, $\epsilon(t)$ accounts for sensor noise and unmodeled disturbances, and $T$ is the number of samples in the observation window. Given paired inputs $\{x(t),u(t)\}$, the objective is to decompose the observation into physiologically interpretable and motion-related components while robustly estimating HR and RR. To this end, we adopt an encoder--decoder architecture in which a motion-conditioned encoder extracts separation-aware representations and a physically constrained decoder reconstructs a structured physiological signal decomposition from these representations.

\subsection{Cross-Modal Conditioning Encoder}
The encoder extracts separation-relevant representations from wrist PPG by jointly processing optical and inertial signals in a dual-stream architecture. This cross-modal conditioning enables adaptive feature modulation across temporal scales and reduces spurious coupling between motion and physiological dynamics.

\subsubsection{PPG Encoder}
The PPG stream is processed by a lightweight residual network tailored for 1D biosignals. The design prioritizes parameter efficiency while preserving morphological detail critical for respiratory analysis. It consists of a stem layer (kernel size 7, stride 2) followed by three residual stages, each comprising two Basic Residual Blocks with 1D convolutions, Batch Normalization, and SiLU activations. Dilated convolutions ($d \in \{1,2,3,4\}$) are employed to capture multiscale temporal features without increasing model complexity, preserving the temporal resolution required to model subtle respiratory modulations.

\subsubsection{Motion-Aware Conditioning via FiLM}
The accelerometer signal $u(t)$ is processed by a lightweight 1D CNN comprising three convolutional blocks with kernel sizes $k \in \{7,5,3\}$ and strided pooling, progressively increasing the channel dimension to $C_{\mathrm{acc}}=64$. Global average pooling produces multi-scale motion embeddings
$\{\mathbf{z}_{\text{early}},\mathbf{z}_{\text{mid}},\mathbf{z}_{\text{late}}\}$,
which condition the corresponding stages of the PPG encoder via Feature-wise Linear Modulation (FiLM)~\cite{perez2018film}.  

Given a PPG feature map $\mathbf{H}$ and a motion embedding vector $\mathbf{z}$, FiLM applies an affine modulation:
\begin{equation}
\mathrm{FiLM}(\mathbf{H}\mid\mathbf{z}) =
\boldsymbol{\gamma}(\mathbf{z}) \odot \mathbf{H}
+ \boldsymbol{\beta}(\mathbf{z}),
\end{equation}
where $\boldsymbol{\gamma}(\mathbf{z})$ and $\boldsymbol{\beta}(\mathbf{z})$ are channel-wise scaling and shifting vectors predicted by learned linear projection heads. This conditioning enables motion-adaptive feature modulation without explicitly subtracting motion-related waveforms.

The resulting motion-conditioned representations serve as inputs to a physically constrained decoder that reconstructs a structured physiological signal through harmonic synthesis.
\subsection{Physically-Constrained Harmonic Decoder}
We employ a \emph{physics-informed harmonic decoder} inspired by Differentiable Digital Signal Processing~\cite{engel2020ddsp} and classical models of PPG signal composition~\cite{Charlton2023}. The decoder is built around an explicit generative model of the physiological PPG signal, enforcing quasi-periodic structure by construction and enabling identifiable signal decomposition.

\subsubsection{Harmonic synthesis}
The clean physiological component is modeled as a harmonic signal with time-varying parameters. Given an instantaneous fundamental frequency trajectory $f_0(t)$, the phase is computed as
\begin{equation}
\theta(t) = 2\pi \sum_{\tau=0}^{t} \frac{f_0(\tau)}{f_s}.
\end{equation}
The physiological signal is synthesized as
\begin{equation}
\mu_{\text{phys}}(t) = (1 + g(t)) \sum_{k=1}^{K} \frac{A_k(t)}{k} \sin(k\theta(t)) + b(t),
\label{eq:synthesis}
\end{equation}
where $A_k(t)$ are the harmonic amplitudes controlling pulse morphology, the gain $g(t)$ models slow amplitude modulation, and the baseline $b(t)$ captures low-frequency variations due to respiration and sensor drift.

\subsubsection{Multi-head parameter prediction}
Motion-conditioned features from the encoder are processed by a shared \emph{refinement block} consisting of two 1D convolutional layers ($C=256$, kernel size $k=3$) with Batch Normalization and SiLU activation. The refined representation is then routed to multiple parallel heads that predict the parameters of the harmonic synthesis model, along with a residual artifact term and a time-varying uncertainty estimate. All convolutional layers use stride $1$ with appropriate padding, ensuring that all predicted trajectories preserve the temporal resolution $T$ of the input PPG signal.

\paragraph{Fundamental frequency}
The cardiac fundamental frequency $f_0(t) \in [0.5, 3.0]$ Hz is predicted using a pointwise 1D convolution and bounded with a sigmoid activation. This constraint stabilizes phase integration and improves identifiability.

\paragraph{Harmonic amplitudes}
Time-varying harmonic amplitudes $A_k(t) \in \mathbb{R}^+$ for $k=1,2,3$ are predicted using a pointwise 1D convolution. Using three harmonics captures the dominant pulse morphology while limiting overfitting to high-frequency noise.

\paragraph{Global gain and baseline}
The global gain $g(t) \in [0,1]$ and baseline term $b(t) \in \mathbb{R}$ are predicted using pointwise 1D convolutional heads. The gain models slow pulse amplitude variations due to contact changes and hemodynamic effects, while the baseline captures low-frequency respiratory effects and sensor drift. A fixed moving-average filter is applied to the baseline to enforce spectral separation from cardiac content.

\paragraph{Residual artifact}
To capture signal components not explained by the harmonic model, a residual artifact $r(t)$ is predicted using a three-layer convolutional head with channel dimensions $\{64,32,1\}$ and kernel size $k=3$. This branch models non-harmonic, motion-induced distortions.

\paragraph{Uncertainty}
The predicted uncertainty reflects local signal unreliability due to irrecoverable motion. A time-varying uncertainty estimate $\sigma^2(t)$ is predicted as the log-variance $\log\sigma^2(t)$ by a two-layer convolutional head. Higher variance indicates segments corrupted by unrecoverable motion and is used to downweight unreliable intervals during reconstruction and inference.

The final reconstructed signal is given by
\begin{equation}
s(t) = \mu_{\text{phys}}(t) + r(t),
\end{equation}
where $\mu_{\text{phys}}(t)$ and $r(t)$ now denote the \emph{predicted} physiological component, obtained from the harmonic synthesis in Eq.~\eqref{eq:synthesis}, and the residual produced by the decoder, respectively. The decoded harmonic parameters are physiologically interpretable: the recovered fundamental frequency directly enables HR estimation, while slower temporal modulations of the amplitudes and frequency encode respiratory dynamics. HR and RR estimation are learned jointly through a multi-objective training formulation that balances reconstruction fidelity, physiological separation, and rate estimation accuracy.

\subsection{Multi-Objective Optimization}

Training employs a composite objective that enforces reconstruction fidelity, identifiable physiological--motion separation, temporal smoothness of decoded parameters, and accurate vital-sign estimation.

\paragraph{Reconstruction and spectral consistency}
Waveform morphology and frequency content are constrained using a negative Pearson correlation loss~\cite{li2014remote} and a multi-resolution spectral loss~\cite{yamamoto2020parallel}:
\begin{align}
\mathcal{L}_{\rho}
&=
1 - \frac{\mathrm{Cov}(x,s)}{\sigma_x \sigma_{s}}, \\[4pt]
\mathcal{L}_{\mathrm{spec}}
&=
\mathbb{E}_m\!\left[
\left\|
\log\!\left|\mathrm{STFT}_m(x)\right|
-
\log\!\left|\mathrm{STFT}_m(s)\right|
\right\|_1
\right],
\end{align}
where $x \in \mathbb{R}^{T}$ is the observed PPG signal, $s \in \mathbb{R}^{T}$ is its reconstruction, $\sigma_x$ and $\sigma_{s}$ denote the corresponding standard deviations, and $\mathrm{STFT}_m(\cdot)$ is computed with window sizes $m \in \{32,64,128,256,512\}$. The reconstruction loss is
\begin{equation}
\mathcal{L}_{\mathrm{rec}} = \mathcal{L}_{\rho} + \mathcal{L}_{\mathrm{spec}} .
\end{equation}

\paragraph{Push--pull separation constraints}
Physiological--motion disentanglement is enforced via
\begin{equation}
\mathcal{L}_{\mathrm{sep}} =
\mathcal{L}_{\mathrm{pull}}
+ \lambda_{\mathrm{orth}}\,\mathcal{L}_{\mathrm{orth}}
+ \mathcal{L}_{\mathrm{noise}} .
\end{equation}

\paragraph*{Physiology pull}
To prevent collapse of the physiological component, we enforce sufficient energy in the cardiac band:
\begin{equation}
\mathcal{L}_{\mathrm{pull}} =
\max\!\left(0, \alpha - P_{\mathrm{band}}(\mu_{\mathrm{phys}})\right),
\end{equation}
where $\mu_{\mathrm{phys}}\in\mathbb{R}^{T}$ is denotes the reconstructed physiological component (pulse-synchronous cardiac signal),
$P_{\mathrm{band}}(\cdot)$ denotes the proportion of power spectral density within
$[f_{\mathrm{HR}}^{\min}, f_{\mathrm{HR}}^{\max}]$, and $\alpha$ is a fixed minimum energy threshold.

\paragraph*{Orthogonality constraint}
Statistical independence between the physiological signal and the residual is encouraged via cosine similarity. For a mini-batch of size $B$,
\begin{equation}
\mathcal{L}_{\mathrm{orth}} =
\frac{1}{B}\sum_{i=1}^{B}
\frac{\left|\boldsymbol{\mu}_{\mathrm{phys}}^{(i)} \cdot \mathbf{r}^{(i)}\right|}
{\|\boldsymbol{\mu}_{\mathrm{phys}}^{(i)}\|_2 \, \|\mathbf{r}^{(i)}\|_2}.
\end{equation}

where $\mathbf{r}^{(i)} \in \mathbb{R}^{T}$ is the residual signal for the $i$-th sample. Orthogonality is enforced as a soft regularization; separation can degrade under non-stationary motion that spectrally overlaps with cardiac dynamics; the complementary spectral and uncertainty-aware terms mitigate the resulting residual leakage.
\paragraph*{Residual noise constraint}
To discourage periodic structure in the residual, cardiac-band energy is penalized:
\begin{equation}
\mathcal{L}_{\mathrm{noise}} = P_{\mathrm{band}}(\mathbf{r}).
\end{equation}
Here, $P_{\mathrm{band}}(\mathbf{r})$ denotes the normalized proportion of spectral power of the residual signal $\mathbf{r}$ contained within the cardiac frequency band $[f_{\mathrm{HR}}^{\min}, f_{\mathrm{HR}}^{\max}]$, computed relative to its total spectral energy.

\paragraph{Parameter smoothing}
Temporal smoothness of decoded physiological parameters is imposed using total-variation regularization:
\begin{equation}
\mathcal{L}_{\mathrm{sm}} =
\mathcal{L}_{\mathrm{diff}}(f_0)
+ \mathcal{L}_{\mathrm{diff}}(g)
+ \sum_{k=1}^{K} \mathcal{L}_{\mathrm{diff}}(A_k),
\end{equation}
with
\begin{equation}
\mathcal{L}_{\mathrm{diff}}(y)=
\frac{1}{T-1}\sum_{t=1}^{T-1}(y_{t+1}-y_t)^2 ,
\end{equation}

where $f_0(t)$, $g(t)$, and $A_k(t)$ denote the predicted fundamental frequency, gain, and harmonic amplitudes.
\paragraph{Heart- and respiratory-rate estimation}
Heart rate is obtained from the recovered fundamental frequency via reliability-weighted temporal averaging,
\begin{equation}
\widehat{\mathrm{HR}} =
60 \cdot
\frac{\sum_{t} w(t)\,f_0(t)}{\sum_{t} w(t)},
\qquad
w(t)=\exp\!\bigl(-\log\sigma^2(t)\bigr),
\end{equation}
where $\sigma^2(t)$ denotes the time-varying uncertainty predicted by the decoder. 

Respiratory rate is predicted from a fused representation of decoded physiological modulations by concatenating the estimated gain (amplitude modulation), baseline variation, fundamental frequency, and harmonic amplitude modulations into a feature vector, which is mapped to a scalar estimate $\widehat{\mathrm{RR}}$ using a lightweight 1D convolutional sub-network and bounded to $[\mathrm{RR}_{\min}, \mathrm{RR}_{\max}]$ via a sigmoid. 
 Direct $\ell_1$ supervision is applied to both vital signs:
\begin{equation}
\mathcal{L}_{\mathrm{hr}}=
\left|\widehat{\mathrm{HR}}-y_{\mathrm{HR}}\right|,
\qquad
\mathcal{L}_{\mathrm{rr}}=
\left|\widehat{\mathrm{RR}}-y_{\mathrm{RR}}\right|,
\end{equation}
where $y_{\mathrm{HR}}$ and $y_{\mathrm{RR}}$ denote the ground-truth heart and respiratory rates, respectively.

\paragraph{Loss weighting}
The total objective combines all losses using homoscedastic uncertainty weighting~\cite{kendall2018multi}:
\begin{equation}
\label{eq:wieghts}
\mathcal{L}_{\mathrm{total}} =
\sum_{\ell \in \{\mathrm{rec},\,\mathrm{sep},\,\mathrm{sm},\,\mathrm{hr},\,\mathrm{rr}\}}
\frac{1}{2\,\tau_{\ell}^{2}}\,\mathcal{L}_{\ell}
+ \log \tau_{\ell}^{2},
\end{equation}
where $\tau_{\ell}^{2} \in \mathbb{R}^{+}$ are learnable scalars that balance the reconstruction, separation, smoothness, HR, and RR losses.

\section{Experiments}
\label{sec:Experiments}

\subsection{Data Acquisition}
\label{sec:datasets}

\textbf{PPG-DaLiA:}
We evaluate our method on the PPG-DaLiA dataset~\cite{reiss2019deepppg}, which is designed to benchmark robust vital-sign estimation from wrist-worn sensors under motion-intensive, free-living conditions. The dataset comprises approximately $36\,\mathrm{h}$ of synchronized recordings from 15 subjects performing eight daily-life activities, including sitting, walking, cycling, stair ascent/descent, driving, and working, with labeled transition periods. Wrist data are collected using an Empatica E4 on the non-dominant wrist, providing PPG sampled at $64\,\mathrm{Hz}$ and tri-axial accelerometer data sampled at $32\,\mathrm{Hz}$. Reference signals are acquired using a chest-worn RespiBAN Professional device, including ECG and respiratory inductive plethysmography sampled at $700\,\mathrm{Hz}$. Ground-truth heart rate is derived from the chest-worn ECG signal, while respiratory rate ground truth is obtained from the chest-worn respiration waveform using peak detection on the clean excursion signal, following standard procedures~\cite{kazemi2025respiration,schafer2008estimation}.

\subsection{Experimental Setup}

\paragraph{Preprocessing}
All signals are segmented into 8\,s windows with a 2\,s stride, following established protocols for wrist PPG analysis~\cite{kechris2024kid,benfenati2025enhanceppg}. PPG and accelerometer signals are temporally aligned and resampled to a common sampling rate of 64 Hz. PPG windows are normalized to zero mean and unit variance, while accelerometer signals are kept in their original scale and used solely for motion conditioning.

\paragraph{Evaluation protocol}
We adopt a leave-one-subject-out (LOSO) evaluation protocol, consistent with prior studies on PPG-DaLiA~\cite{reiss2019deepppg,kechris2024kid}. In each fold, data from one subject are held out for testing while the remaining subjects are used for training. Performance is reported as the mean across all subjects using mean absolute error (MAE).

For RR estimation, the proposed model produces an RR prediction for each 8\,s window. To enable fair comparison with prior methods that rely on longer temporal contexts (e.g., 32\,s windows~\cite{kazemi2025respiration}), we report results for both the raw 8\,s predictions, reflecting low-latency estimation, and aggregated 32\,s estimates obtained by averaging four consecutive windows.

\paragraph{Training details}
All models are implemented in PyTorch and trained end-to-end using the AdamW optimizer with a learning rate of $6\times10^{-4}$ and weight decay of $10^{-4}$. Training is conducted for 60 epochs with a batch size of 64, using cosine annealing for learning-rate scheduling. The proposed PCHS model contains approximately 1.97M trainable parameters, comparable to recent PPG-based deep learning approaches.

\subsection{Experimental Results}
\label{sec:results}
To better understand the contribution of each design component, we conduct ablation studies to isolate the effect of the key architectural and modeling choices in the proposed framework.

\subsubsection{Ablation Study}

Table~\ref{tab:ablation} quantifies the contribution of the key components of the proposed Physically-Constrained Harmonic Separation (PCHS) framework through targeted ablations. We evaluate the impact of (i) motion-aware conditioning via FiLM, (ii) explicit residual modeling with separation regularization, (iii) physics-guided harmonic synthesis, and (iv) uncertainty-aware inference.

\paragraph{Motion-aware conditioning}
Removing FiLM-based conditioning prevents accelerometer-derived motion context from modulating physiological features, leading to a clear degradation in performance (HR MAE increases from 3.20 to 3.75 bpm; $-17.2\%$). This confirms that explicit motion conditioning enables adaptive artifact suppression at the feature level, beyond what implicit learning alone can achieve.

\paragraph{Residual modeling and separation}
In the full model, the signal is decomposed as $x(t)=\mu_{\mathrm{phys}}(t)+r(t)$, with a separation loss enforcing orthogonality and suppressing cardiac-band energy in the residual. Removing the residual pathway yields the largest performance drop (HR MAE 4.10 bpm; $-28.1\%$), indicating that forcing the harmonic model to absorb broadband noise compromises identifiability. Disabling only the separation constraints also degrades accuracy (HR MAE 3.70 bpm), highlighting the necessity of explicit push--pull regularization to prevent physiological leakage.

\paragraph{Physics-guided harmonic synthesis}
Replacing the harmonic analysis-by-synthesis generator with direct waveform regression significantly reduces accuracy (HR MAE 3.92 bpm; $-22.5\%$). This demonstrates that enforcing quasi-periodicity and smooth amplitude modulation through a harmonic prior is critical for stable frequency estimation in low-SNR, motion-corrupted conditions.

\paragraph{Uncertainty-aware inference}
Disabling reliability-guided temporal aggregation results in a consistent performance drop (HR MAE 3.55 bpm; $-10.9\%$). By down-weighting segments with high predicted uncertainty, the model mitigates transient corruption that cannot be resolved by structural constraints alone.

Overall, the ablation results show that PCHS derives its robustness from the complementary interaction of motion conditioning, physics-guided decomposition, explicit residual modeling, and uncertainty-aware inference, rather than from any single component in isolation.

\begin{table}[t]
\centering
\caption{Ablation study on PPG-DaLiA. Accuracy change denotes relative performance drop compared to the full PCHS model.}
\label{tab:ablation}
\begin{tabular}{lcc}
\hline
\textbf{Variant} & \textbf{HR MAE} $\downarrow$ (bpm) & \textbf{Accuracy Drop} (\%) \\
\hline
\textbf{Ours (Full PCHS)} & \textbf{3.20} & -- \\
w/o FiLM conditioning & 3.75 & $-17.2$ \\
w/o Residual pathway & 4.10 & $-28.1$ \\
w/o Separation loss & 3.70 & $-15.6$ \\
w/o Harmonic synthesis & 3.92 & $-22.5$ \\
w/o Reliability weighting & 3.55 & $-10.9$ \\
\hline
\end{tabular}
\vspace{-0.5cm}
\end{table}

\paragraph{Adaptive Loss Balancing Analysis}
\label{sec:ablation_weighting}

As part of the ablation analysis, we examine the effect of adaptive loss balancing using homoscedastic uncertainty weighting~\cite{kendall2018multi}, as defined in eq.\ref{eq:wieghts} Sec.~\ref{sec:Methods}, which allows the model to dynamically prioritize tasks based on their estimated uncertainty.

Table~\ref{tab:weights_evolution} illustrates the evolution of the learned loss weights during training. Once the physics-constrained decoder stabilizes, a clear reallocation of task importance emerges. In particular, the Respiratory Rate objective receives the highest final weight ($\lambda \approx 1.32$), indicating reduced aleatoric uncertainty when respiratory features are extracted from the recovered physiological signal $\mu_{\mathrm{phys}}$. In contrast, the separation loss ($\mathcal{L}_{\mathrm{sep}}$ remains relatively low-weighted ($\lambda \approx 0.20$), reflecting its role as a regularizer rather than a primary optimization target. This adaptive weighting enables the model to prioritize clinically relevant rate estimation without destabilizing the learning of harmonic representations.

\begin{table}[htbp]
\centering
\caption{Evolution of learned loss weights ($\lambda$) during training. \textbf{Bold} indicates the highest final weight.}

\label{tab:weights_evolution}
\resizebox{\columnwidth}{!}{%
\begin{tabular}{lcccc}
\hline
\textbf{Loss Term} & \textbf{Epoch 1} & \textbf{Epoch 10} & \textbf{Epoch 20} & \textbf{Ep 30} \\ \hline
Reconstruction ($\mathcal{L}_{rec}$) & 0.10 & 0.22 & 0.35 & 0.41 \\
Separation ($\mathcal{L}_{sep}$) & 0.10 & 0.15 & 0.18 & 0.20 \\
Smoothness ($\mathcal{L}_{sm}$) & 0.10 & 0.25 & 0.38 & 0.45 \\
Heart Rate ($\mathcal{L}_{hr}$) & 0.10 & 0.55 & 0.82 & 0.95 \\
Respiratory Rate ($\mathcal{L}_{rr}$) & 0.10 & 0.60 & 1.15 & \textbf{1.32} \\ \hline
\end{tabular}%
}
\vspace{-0.2cm}
\end{table}

\subsubsection{Comparison with the state-of-the-art}

\paragraph{Heart Rate estimation}
Table~\ref{tab:ppgdalia_per_subject} shows that the proposed PCHS framework achieves the lowest average HR mean absolute error on the PPG-DaLiA dataset ($3.20\pm1.21$ bpm), outperforming state-of-the-art signal-processing and deep learning approaches across all 15 subjects. Performance gains are consistent under both low- and high-motion conditions, with particularly strong improvements for subjects affected by severe motion (e.g., S5, S8, and S9).

In comparison, existing signal-processing and learning-based methods exhibit reduced robustness under strong motion, reflecting the challenges posed by non-stationary and spectrally overlapping artifacts in free-living conditions. While recent deep learning and knowledge-informed approaches improve overall accuracy, variability remains for heavily corrupted recordings.

By treating motion as a conditioning signal rather than a predictive shortcut, PCHS enables motion-aware physiological inference without suppressing pulse-synchronous cardiac content. This physiology-driven analysis-by-synthesis formulation isolates cardiac structure from motion-related corruption, yielding interpretable and identifiable representations that support robust heart-rate estimation.

\begin{table*}[!thb]
\centering
\caption{Per-subject HR MAE (bpm) on PPG-DaLiA. Best results are highlighted in \textbf{bold}.}
\label{tab:ppgdalia_per_subject}
\resizebox{\textwidth}{!}{
\begin{tabular}{lcccccccccccccccc}
\hline
Method & S1 & S2 & S3 & S4 & S5 & S6 & S7 & S8 & S9 & S10 & S11 & S12 & S13 & S14 & S15 & Mean \\
\hline
\multicolumn{17}{l}{\textbf{Signal Processing}} \\
\hline
SpaMaPlus~\cite{reiss2019deepppg}  (PPG,ACC)
& 8.86 & 9.67 & 6.40 & 14.10 & 24.06 & 11.34 & 6.31 & 11.25 & 16.04 & 6.17 & 15.15 & 12.03 & 8.50 & 7.76 & 8.29 & 11.06 \\
TAPIR~\cite{huang2020robust} (PPG)
& 4.50 & 4.50 & 3.20 & 6.00 & 5.00 & 3.40 & 2.80 & 6.30 & 8.00 & 2.90 & 5.10 & 4.70 & 3.10 & 5.00 & 4.10 & 4.57 \\
CurToSS~\cite{zhou2020heart} (PPG)
& 5.40 & 4.30 & 3.00 & 8.00 & 2.20 & 2.80 & 3.30 & 8.50 & 12.60 & 3.60 & 3.60 & 6.10 & 3.00 & 5.50 & 3.70 & 5.04 \\
\hline
\multicolumn{17}{l}{\textbf{Deep Learning}} \\
\hline
DeepPPG~\cite{reiss2019deepppg}(PPG,ACC)
& 7.73 & 6.74 & 4.03 & 5.90 & 18.51 & 12.88 & 3.91 & 10.87 & 8.79 & 4.03 & 9.22 & 9.35 & 4.29 & 4.37 & 4.17 & 7.65 \\

TEMPONet~\cite{burrello2022improving}(PPG,ACC)
& 4.37 & 3.74 & 2.43 & 5.49 & 13.48 & 5.71 & 2.23 & 7.86 & 8.94 & 3.32 & 5.34 & 7.71 & 2.03 & 2.94 & 3.58 & 5.27 \\

AugmentPPG~\cite{burrello2022improving}(PPG,ACC)
& 4.37 & 3.74 & 2.43 & 5.49 & 9.41 & 3.63 & 2.23 & 7.86 & 8.94 & 3.32 & 5.34 & 7.64 & 2.03 & 2.94 & 3.58 & 4.86 \\
PULSE~\cite{kasnesis2023feature}(PPG,ACC)
& 3.78 & 3.04 & 2.20 & 4.41 & 6.95 & 3.71 & 2.39 & 8.17 & 6.19 & 2.60 & 3.85 & 5.22 & 1.98 & 3.13 & 2.79 & 4.03 \\
KID-PPG~\cite{kechris2024kid}(PPG,ACC)
& 4.27 & 3.46 & 2.07 & 5.61 & 3.01 & 2.74 & 1.39 & 7.13 & 9.53 & 2.77 & 3.58 & 4.52 & 1.48 & 2.48 & 2.84 & 3.79 \\
EnhancePPG~\cite{benfenati2025enhanceppg}(PPG,ACC)
& 3.35 & 3.15 & 2.20 & 4.38 & 5.64 & 2.35 & 1.93 & 5.16 & 6.38 & 2.87 & 3.30 & 5.49 & 1.84 & 2.43 & 2.53 & 3.54 \\

\hline
PCHS(OURS) (PPG,ACC)
& 3.52 & 2.92 & 2.40 & 5.41 & 3.23
& 3.52 & 1.75 & 5.07 & 5.65 & 2.07
& 3.13 & 3.60 & 1.24 & 2.20 & 2.27& \textbf{3.20} \\
\hline

\end{tabular}}
\end{table*}

\paragraph{Respiratory-rate estimation on PPG-DaLiA}
Table~\ref{tab:comparison_dalia_avg} compares the proposed PCHS model with 
recent state-of-the-art methods for RR estimation on PPG-DaLiA. 
Because RR has a low-frequency content (periods of roughly 2--5\,s), 
reliable estimation favors longer temporal context spanning multiple 
respiratory cycles, which motivates evaluation at two operating points.
At \textbf{Raw 8\,s} (low-latency inference), PCHS achieves an MAE of 
3.54 bpm, competitive with deep learning baselines such as 
RRWaveNet~\cite{osathitporn2023rrwavenet} (3.41 bpm) and substantially 
better than hybrid heuristic approaches like A--P Synthesis~\cite{huang2021novel}
(8.12 bpm). At \textbf{Agg.\ 32\,s}, matching the temporal resolution of 
prior work, PCHS reaches \textbf{2.15 bpm}, outperforming the 
transfer-learning-based approach of~\cite{kazemi2025respiration} (2.29 bpm) 
and the MultiScale CNN of~\cite{kazemi2023robust} (2.70 bpm).

\begin{table}[!htb]
\centering
\caption{Comparison of Average RR MAE (bpm) on the PPG-DaLiA dataset. Best results are highlighted in \textbf{bold}.}
\label{tab:comparison_dalia_avg}
\begin{tabular}{llc}
\hline
\textbf{Method} & \textbf{Input} & \textbf{Avg MAE} \\ \hline
CNN \cite{bian2020respiratory} & PPG & 3.16 \\
A-P Synthesis \cite{huang2021novel} & PPG, ACC & 8.12 \\
MultiScale CNN \cite{kazemi2023robust} & PPG, ACC, Gyr & 2.70 \\
RRWaveNet \cite{osathitporn2023rrwavenet} & PPG & 3.41 \\
Transfer Learning \cite{kazemi2025respiration} & PPG, ACC & 2.29 \\ \hline
\textbf{PCHS(OURS) (Raw 8\,s)} & PPG, ACC & 3.54 \\
\textbf{PCHS(OURS) (Agg. 32\,s)} & PPG, ACC & \textbf{2.15} \\ \hline
\end{tabular}
\vspace{-0.5cm}
\end{table}

\subsubsection{Qualitative interpretability}
\begin{figure}[!htb]
    \centering
    \includegraphics[width=0.9\linewidth]{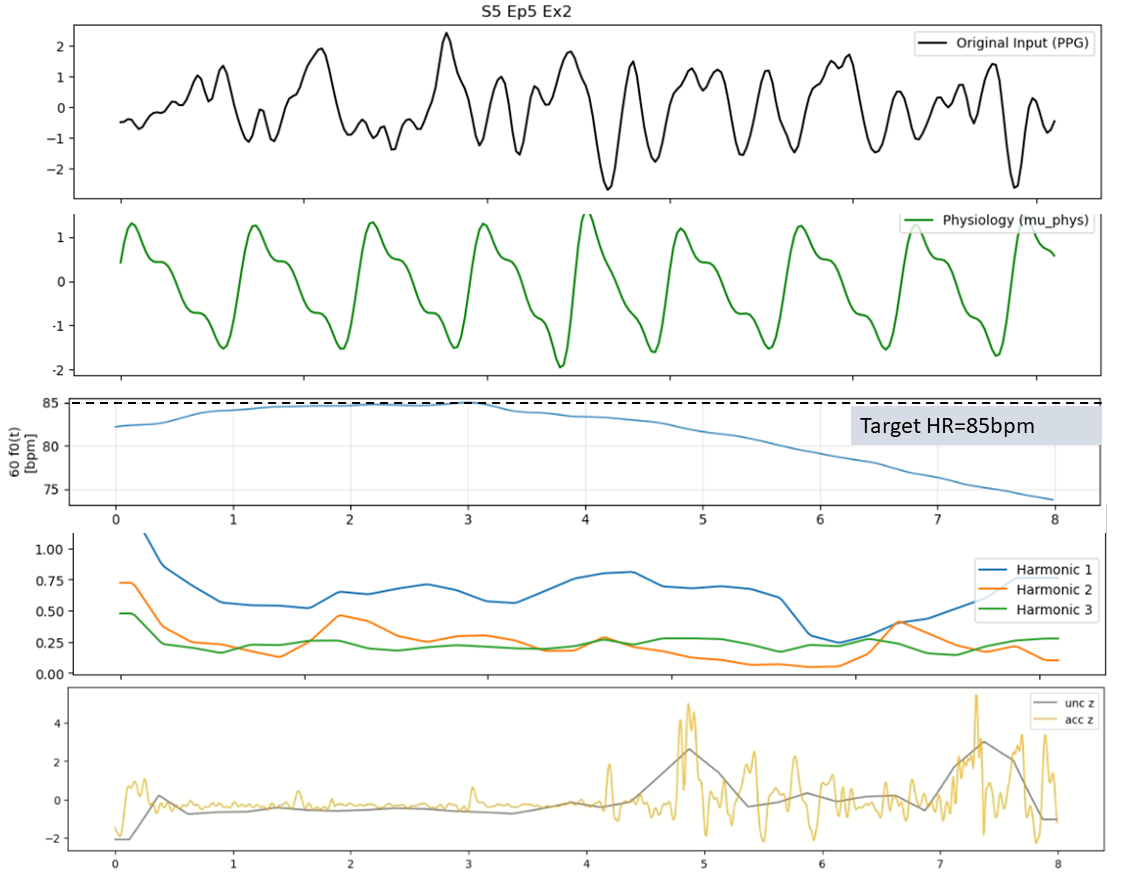}
    \caption{Qualitative visualization of the PCHS decomposition for a representative window from subject S05, showing the raw PPG signal, reconstructed physiological component $\mu_{\mathrm{phys}}(t)$, fundamental frequency, harmonic amplitudes $\{A_k(t)\}$, and motion-related indicators.}

    \label{fig:qualitative}
    
\end{figure}

Beyond quantitative HR/RR accuracy, PCHS provides interpretable intermediate variables for qualitative inspection of the learned decomposition. The decoder explicitly separates the physiologically structured component $\mu_{\mathrm{phys}}(t)$, synthesized from the predicted parameters, from the additive residual $r(t)$ capturing motion-related distortions. Figure~\ref{fig:qualitative} shows a representative example from subject S05, including the raw PPG signal, reconstructed physiological component, and harmonic evolution.
The reconstructed harmonic structure remains smooth and exhibits a stable amplitude hierarchy, indicating that cardiac morphology is explained through the physics-constrained analysis-by-synthesis pathway. Periods of elevated accelerometer activity coincide with increased residual energy and higher predicted uncertainty, consistent with the intended artifact modeling and reliability-aware inference. In contrast, the estimated fundamental frequency remains stable and physiologically plausible throughout the window, demonstrating robust cardiac frequency tracking under motion. Together, these observations support the interpretability and physiological validity of the proposed PCHS framework.

\section{Conclusion}
\label{sec:Conclusion}
This work presented a Physically-Constrained Harmonic Separation framework for robust heart-rate and respiratory-rate estimation from wrist photoplethysmography under motion. By casting vital-sign estimation as an analysis-by-synthesis problem, the proposed approach explicitly decomposes the observed PPG into an interpretable physiological component and a motion-driven residual, treating accelerometer measurements as explanatory cues for corruption rather than predictive shortcuts. A physics-guided harmonic pulse model enables direct recovery of the cardiac fundamental frequency, while slow envelope and baseline modulations capture respiratory dynamics. A time-varying reliability signal further improves robustness by downweighting corrupted intervals during inference. Experiments on the PPG-DaLiA dataset demonstrate improved performance under motion-heavy conditions and yield interpretable decompositions that disentangle pulse structure from motion-induced corruption.
Future work will extend the framework to additional datasets and downstream 
tasks, and investigate efficient deployment strategies for resource-constrained 
wearable devices. Validation on arrhythmic populations, where the 
quasi-periodicity assumption may not hold, is a particularly important 
open direction. More broadly, this study highlights the value of combining physiological structure with learning-based models to achieve robust and interpretable vital-sign estimation in unconstrained wearable sensing scenarios.

\end{document}